\documentclass[traditabstract]{aa} 
\usepackage{graphicx}
\usepackage{txfonts}
\begin{document}
\title{Molecular excitations: a new way to detect Dark matter}

\author{J.Va'vra
\inst{}
}

\institute{SLAC, Stanford University, CA94309, U.S.A.\\
\email{jjv@slac.stanford.edu}
}

\date{Received February 3, 2014}


\abstract
{We believe that the Dark Matter (DM) search should be expanded into the domain of detectors sensitive to molecular excitations, and so that we should create detectors 
which are more sensitive to collisions with very light WIMPs. In this paper we investigate in detail di-atomic molecules, such as Fused Silica material with large 
OH-molecule content, and water molecules. Presently, we do not have suitable low cost IR detectors to observe single photons, however some OH-molecular excitations 
extend to visible and UV wavelengths, and can be measured by Bialkali photocathodes. There are many other chemical substances with di-atomic molecules, or more complex oil 
molecules, which could be investigated also. This idea invites searches in experiments having large target volumes of such materials coupled to a large array of single-photon 
detectors with Bialkali or infrared-sensitive photocathodes.} 


\maketitle
%

\section{Introduction}

\begin{figure}[tbp]
\includegraphics[width=0.45\textwidth]{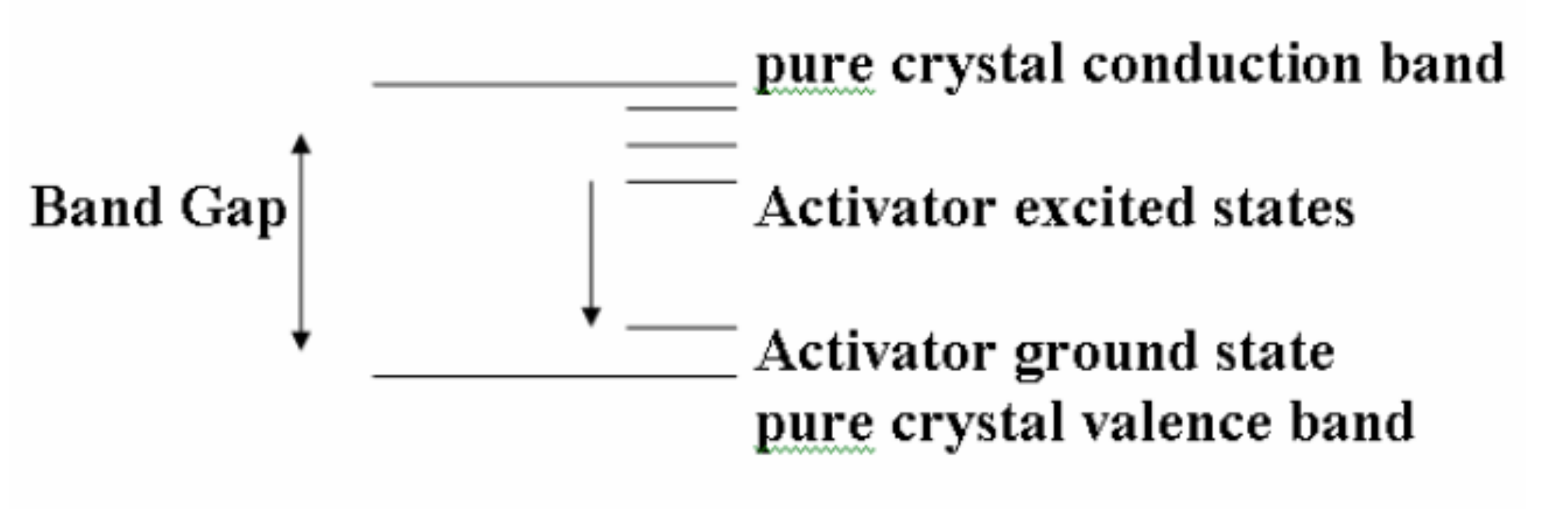}
\caption{Energy levels for an activated NaI(Tl) crystal, where Tl is the so-called activator. The Tl-levels lie somewhat below the conduction band 
so that the radiated photons are emitted with less energy than the band gap of the pure crystal.}
\label{fig:NaI_energy_levels}
\end{figure}

\begin{figure}[tbp]
\includegraphics[width=0.4\textwidth]{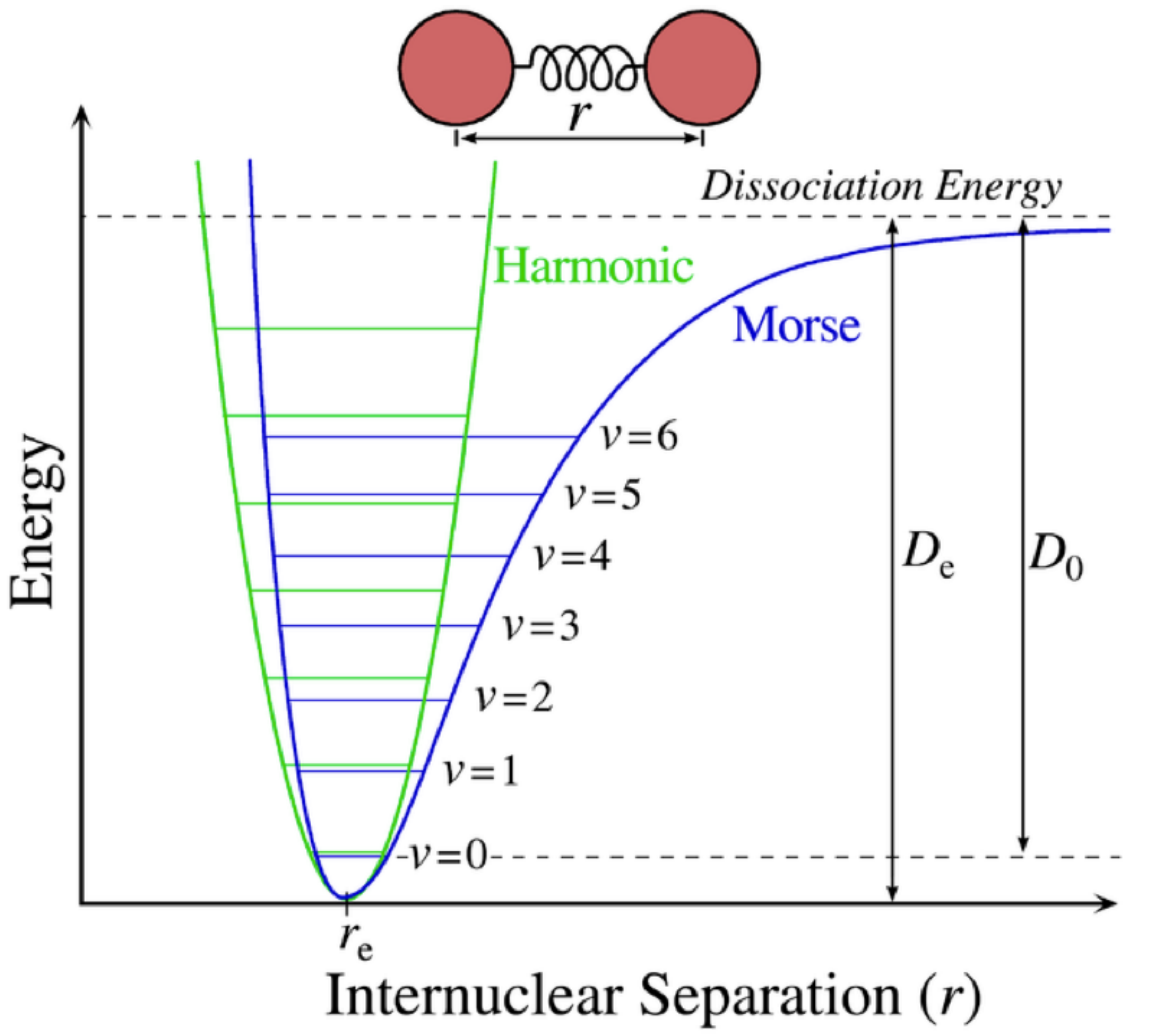}
\caption{Energy levels of diatomic molecules for a harmonic and an anharmonic oscillator, where the latter follows the more complex Morse potential; 
$\nu$ is the vibrational quantum number ($\nu$ = 0,1,2,...). The $\nu$ = 0 level is the vibrational ground state. Unlike 
the energy levels of the harmonic oscillator potential, which are evenly spaced by $\hbar\omega$, the Morse potential level spacing decreases as the energy 
approaches the dissociation energy $D_{e}$ [\cite{Morse_2014}].}
\label{fig:Vibration_spectra_of_diatomic_molecules}
\end{figure}

A recent paper [\cite{Vavra_2014_a}] discussed the possibility that the DAMA experiment [\cite{Bernabei_2013}] observes the effect of a light WIMP striking a hydrogen atom and creating a 
proton recoil, which then excites the NaI(Tl) crystal in which it is located. The target hydrogen atom is present as OH or H$_2$O contamination. The "classical" scintillation mechanism for 
atomic level excitation in NaI(Tl) is shown schematically in Fig.~\ref{fig:NaI_energy_levels}. The primary energy deposit has to exceed the band gap of 
NaI, which is ~3.9 eV; thallium is added to lower the emission energy. 

The present paper considers another, even more sensitive, method to detect Dark Matter via molecular excitations (a preliminary 
version was presented in Ref.~[\cite{Vavra_2014_b}]). A very small energy-deposit, 
at a level of 2-3~eV, is needed to excite such vibrations; this is a factor of 2 lower in energy than the classical scintillation mechanism. Many di-atomic 
molecules will vibrate when hit by a photon (or neutron or a WIMP) and some small fraction of such single photon excitations can be detected by a PMT working in 
the Bialkali regime. However, most of the emitted energy is in the IR wavelength region, or is in the form of heat; the emission of visible photons is suppressed 
by at least 4-6 orders of magnitude compared to IR photons. Because the efficiency for Dark Matter to generate visible photons in water or ice is expected to be 
very small, a large detector volume is required, together with an extensive PMT coverage with single photon sensitivity. Examples of 
possible experiments to do this measurement are the Ice Cube, BaBar DIRC and the Super-Kamiokande experiment.

\begin{table}
\caption{A simple calculation of the transition wavelength for several frequency overtones of the OH-radicals. The last two modes correspond to visible wavelengths. 
Higher modes can reach the UV regime [\cite{Chemistry_lectures}].}
\label{table:1}                     
\resizebox{9cm}{!} {
\begin{tabular}{c c c}                  
\hline\hline                            
OH-band identity  & Transition &  Calculated wavelength [nm]  \\     
\hline                                  
$\nu_{1}$ & 0 $\rightarrow$ 1 & 2803 \\
2$\nu_{1}$ & 0 $\rightarrow$ 2 & 1436 \\
3$\nu_{1}$ & 0 $\rightarrow$ 3 & 980 \\
4$\nu_{1}$ & 0 $\rightarrow$ 4 & 755 \\
5$\nu_{1}$ & 0 $\rightarrow$ 5 & 619.5 \\
\hline                                   
\end{tabular}
}
\end{table}

Di-atomic molecules are ideal for such studies, since they are simpler to understand, and have been 
studied extensively by chemists [\cite{Morse_1929}]. The energy levels of di-atomic molecules can be calculated using the Schroedinger equation 
description of a harmonic oscillator, which has been
described in many textbooks - $e.g.$ see Refs. [\cite{Morse_2014}] and [\cite{Chemistry_lectures}]. 
Figure~\ref{fig:Vibration_spectra_of_diatomic_molecules} and Table~\ref{table:1} provide 
examples of energy levels in a typical di-atomic molecule, for a perfect harmonic oscillator, and for an anharmonic oscillator following the Morse potential [\cite{Morse_2014}]. 
The most significant difference between the two potentials is a transition selection rule. The harmonic oscillator allows only transitions obeying $\bigtriangleup\nu=\pm1$. 
In contrast, the anharmonic oscillator allows the transitions $\bigtriangleup\nu=\pm1$, $\pm2$, $\pm3$, etc. These transitions, 
$\nu_{1}$, 2$\nu_{1}$, 3$\nu_{1}$, etc., are called overtones. 
We see that although the fundamental mode corresponds to an IR wavelength, the higher overtones correspond to visible or even UV wavelengths. This can 
provide a path to detection using Bialkali photocathode-based PMTs; single-photon IR-sensitive detectors are more difficult to utilize at present for this purpose. 

\begin{figure}[tbp]
\includegraphics[width=0.4\textwidth]{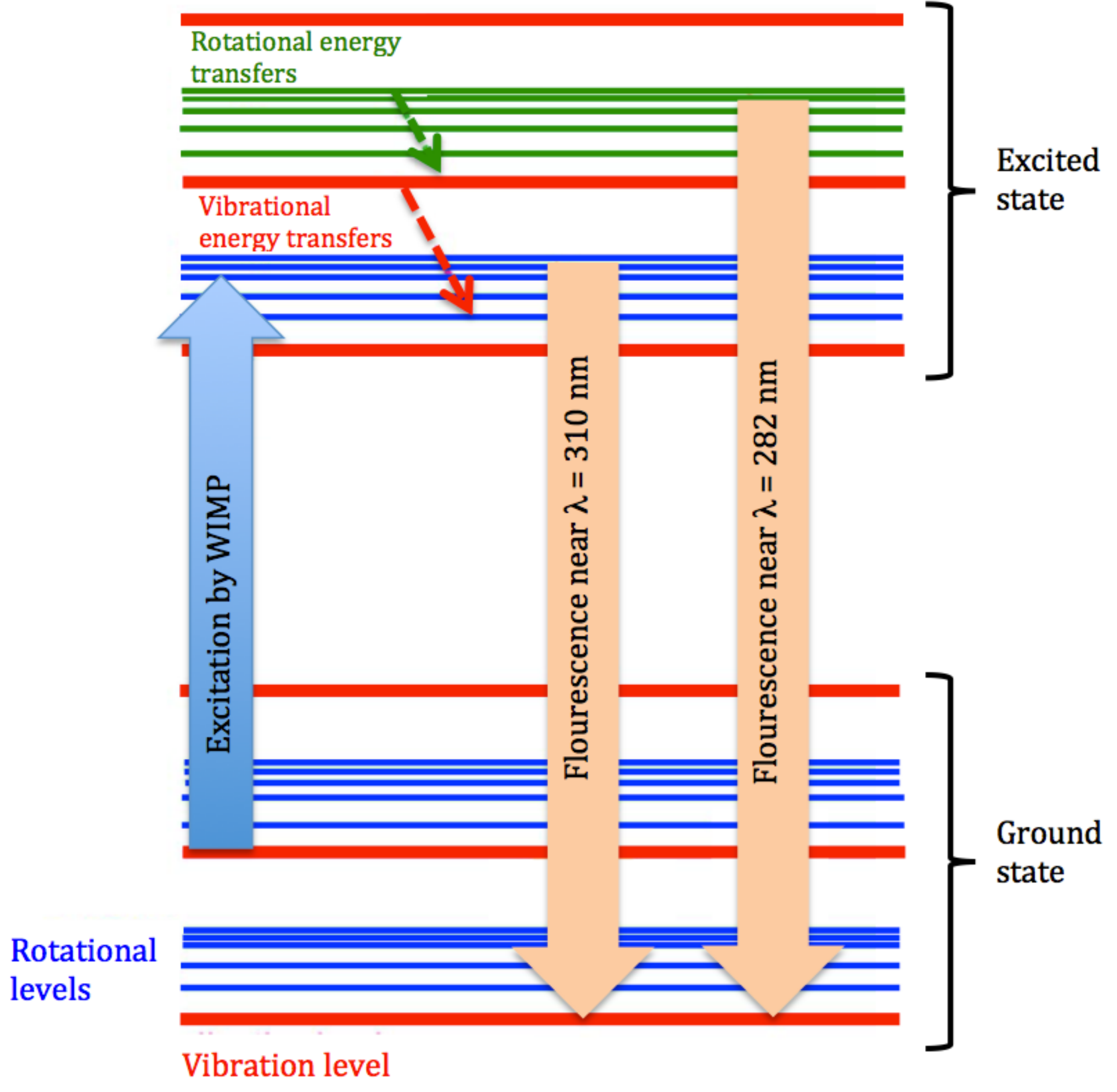}
\caption{A schematic picture of WIMP-scattering excitation of energy levels in the OH-molecule, with subsequent de-excitation 
via fluorescent photons at either $\sim$282~nm or $\sim$310~nm.}
\label{fig:OH_levels}
\end{figure}

We note that the OH-molecule has been studied extensively by many chemists using laser-induced fluorescence. An example of such a fluorescence measurement is
the OH-molecule excitation by a 282~nm dye laser, followed by the observation of the 310~nm wavelength with a PMT with a notch filter [\cite{Smith_1990}]. 
This method was used to determine traces of OH-radicals in the atmosphere [\cite{Matsumi_2002}]. Figure~\ref{fig:OH_levels} shows schematically how WIMP scattering could excite 
energy levels in the OH-molecule, with a subsequent de-excitation via fluorescent photons at either $\sim$282~nm or $\sim$310~nm.

However, the picture outlined above is too simplistic, since the actual energy levels can be affected by the substance in which the OH-radical is located. In this paper we 
discuss the example of Fused Silica loaded with OH-molecules. Fused Silica is a good example because it has been investigated previously in great detail. 
It is an ideal material in which to study the effects of the OH-molecular presence, because the basic material is very pure. In addition, Fused Silica can be 
prepared in the form of long fibers, which allows the accurate measurement of absorption resonances by using a monochromator. 
The next chapter will deal with this material in more detail.

Although Fused Silica is a well-understood material, pure water is even better understood. We have looked into its molecular vibration response also.

\section{OH-molecular vibrations in Fused Silica}

\begin{figure}[tbp]
\includegraphics[width=0.45\textwidth]{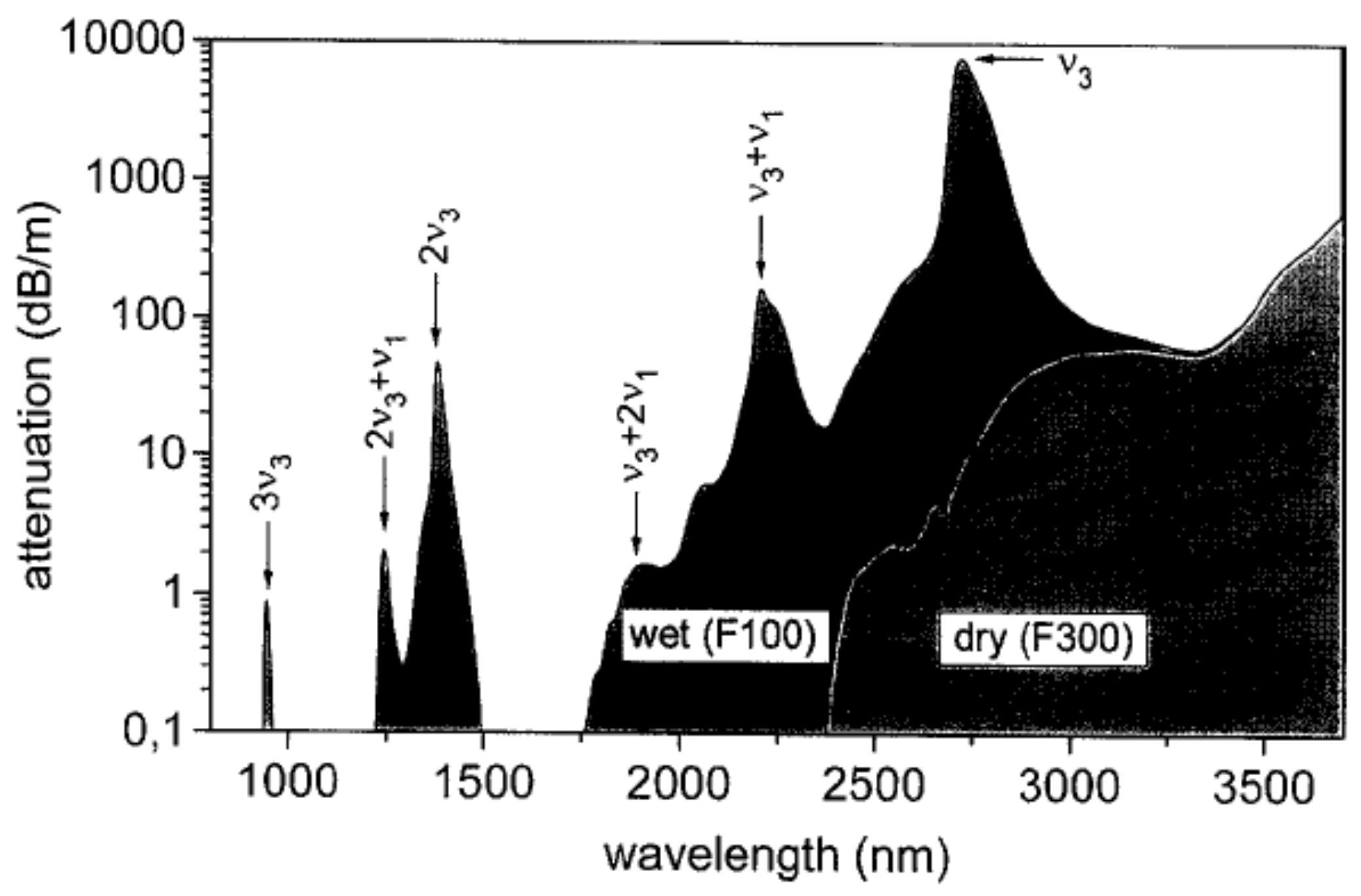}
\caption{Attenuation spectra of wet (F100) and dry (F300) Fused Silica short-bulk samples. Each sample is too short (20~cm) to be sensitive to higher harmonic modes [\cite{Humbach_1996}].}
\label{fig:Quartz_absorption_peaks_bulk}
\end{figure}

Reference [\cite{Humbach_1996}] investigated IR absorption in wet and dry Fused Silica. Wet Fused Silica contains a large quantity of OH-molecules.
In contrast, the dry Fused Silica has almost no OH-content and a minimum of other impurities, so that it is useful for fiber applications operating in the IR regime. 
The OH content is due to the presence of some moisture or hydrogen in the manufacturing process, which results in hydroxyl groups chemically bonded to the silica molecular network (SiOH). 
This hydroxyl (OH) affects the optical properties of silica due to the fundamental OH-absorption band with $\nu_{3}$~=~2.72~$\mu$m, the corresponding overtones,
and the combination modes with the SiO$_{4}$ tetrahedron vibration. The OH content of the dry Fused Silica was 0.2~ppm (designated as F300), and the OH-content of the 
wet Fused Silica was about 700~ppm (designated as F100). Figure~\ref{fig:Quartz_absorption_peaks_bulk} shows the absorption measurements for both types of Fused Silica in short-bulk samples. 
The spectrum of the wet silica is affected by the fundamental mode $\nu_{3}$, the overtones 2$\nu_{3}$, 3$\nu_{3}$, and a few combination modes. Higher 
overtones, n$\nu_{3}$, were not observed because the sample was only 20~cm in length. However, the fundamental $\nu_{3}$ frequency was observed in the wet sample. 
The dry sample does not exhibit major absorption peaks.

\begin{figure}[tbp]
\includegraphics[width=0.45\textwidth]{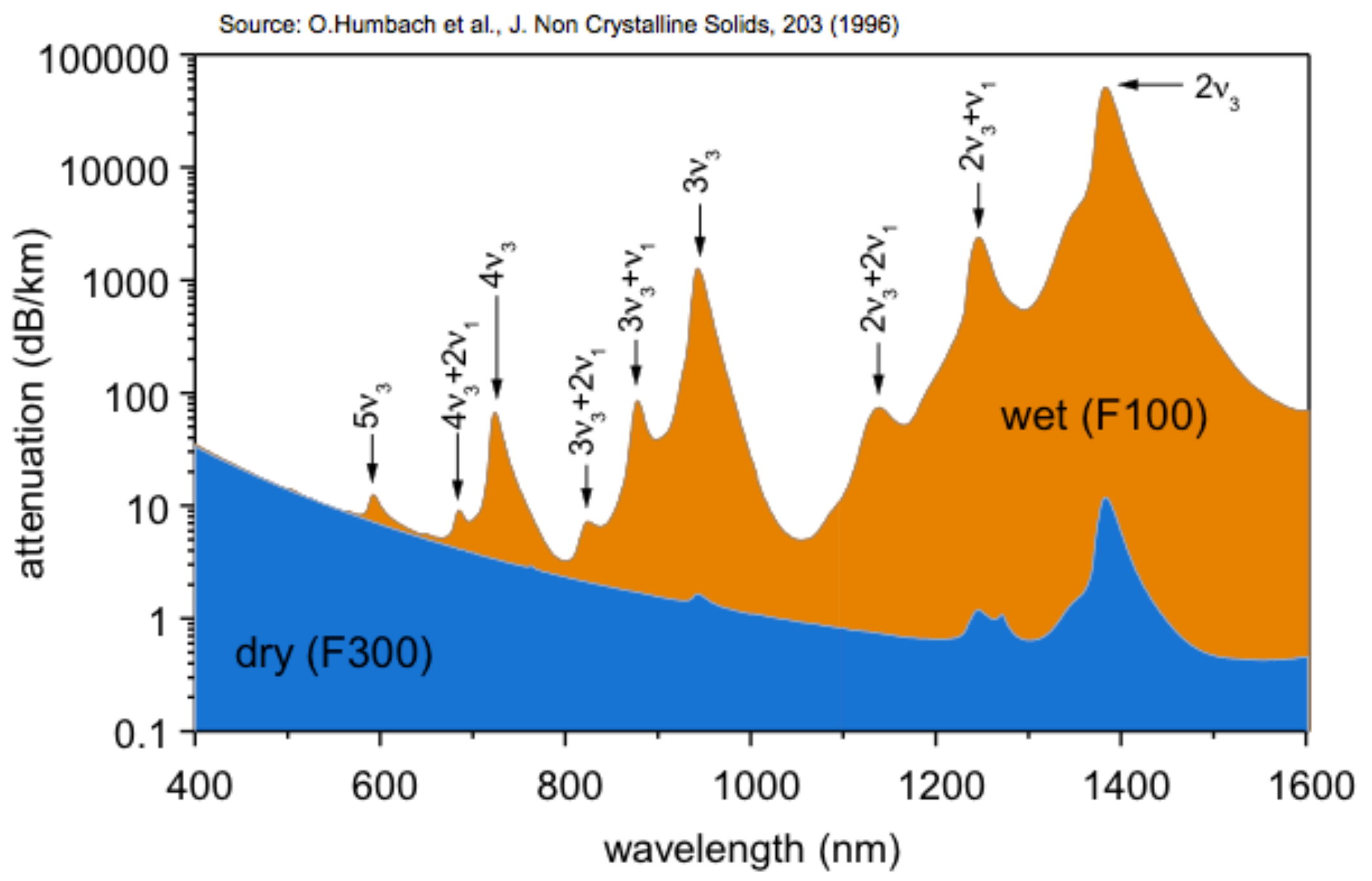}
\caption{Attenuation spectra of wet (F100) and dry (F300) silica core fibers, each about 1~km-long. At lower wavelength the attenuation is dominated by 
Rayleigh scattering, which probably prevents the measurement of even higher overtone modes.
The parameter $\nu_{3}$ corresponds to the OH-molecule fundamental mode, and the parameter $\nu_{1}$ to the SiO$_{4}$ tetrahedron vibration fundamental mode [\cite{Humbach_1996}].}
\label{fig:Quartz_absorption_peaks_fiber_a}
\end{figure}

Figure~\ref{fig:Quartz_absorption_peaks_fiber_a} shows the absorption fiber spectra in the lower wavelength region using fibers 1km in length. The wet silica 
spectrum is a good example of the principal pattern spectral positions and relative intensities of the OH-bands. Each of the overtones at 724 nm, 943 nm 
and 1383 nm is accompanied by two less-intense combination modes located on the short wavelength tail of the corresponding overtone absorption band.
The basic attenuation is almost entirely governed by Rayleigh scattering. This is the reason why the experiment measures only the seventh overtone 7$\nu_{3}$.

\begin{table}
\caption{Measured OH absorption bands in wet Fused Silica (F100), starting with the fundamental mode, $\nu_{3}$, and going all the way up to the 
seventh OH-overtone at 7$\nu_{3}$~=~444 nm. The error on the wavelength measurement is $\pm$1nm [\cite{Humbach_1996}].}
\label{table:2}                                     
\resizebox{7cm}{!} {
\begin{tabular}{c c}                                
\hline\hline                                        
OH-band identity  & Measured wavelength [nm]  \\    
\hline                                              
$\nu_{3}$            & 2722 \\
$\nu_{3}$+$\nu_{1}$  & 2212 \\
$\nu_{3}$+2$\nu_{1}$ & 1894 \\
2$\nu_{3}$           & 1383 \\
2$\nu_{3}$+$\nu_{1}$ & 1246 \\
2$\nu_{3}$+2$\nu_{1}$ & 1139 \\
3$\nu_{3}$           & 943 \\
3$\nu_{3}$+2$\nu_{1}$ & 825 \\
4$\nu_{3}$           & 724 \\
4$\nu_{3}$+$\nu_{1}$ & 685 \\
4$\nu_{3}$+2$\nu_{1}$ & 651 \\
5$\nu_{3}$           & 593 \\
5$\nu_{3}$+$\nu_{1}$ & 566 \\
6$\nu_{3}$           & 506 \\
7$\nu_{3}$           & 444 \\
\hline                                              
\end{tabular}
}
\end{table}
  
Table~\ref{table:2} shows the measured OH-absorption bands in wet Fused Silica (F100), starting with the fundamental mode $\nu_{3}$ and going all the way up to the 
seventh OH-overtone at 7$\nu_{3}$~=~444~nm. The Table also shows combination modes with the SiO$_{4}$ tetrahedron vibration $\nu_{1}$. We note that a collision 
between a proton or an oxygen atom of the OH-molecule and a WIMP could excite a number of vibrations detectable by a Bialkali photocathode. We conclude that the following 
wavelengths fall in the reach of a Bialkali photocathode: 724, 685, 651, 593, 566, 506, and 444~nm.

\section{OH-molecular vibrations in Pure Water and Ice}

\begin{figure}[tbp]
\includegraphics[width=0.4\textwidth]{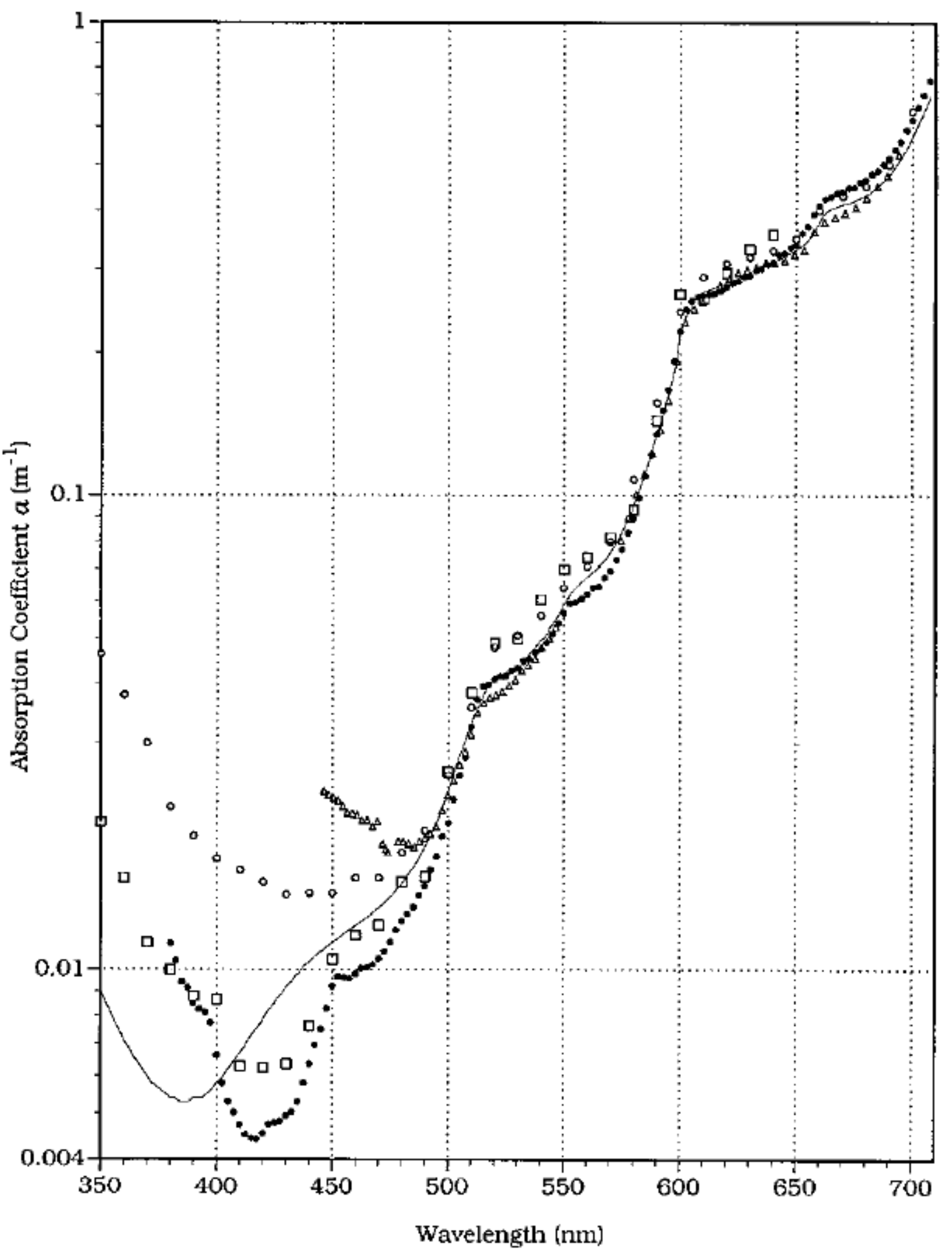}
\caption{Comparison of various measurements of absorption in pure water in the visible wavelength region [\cite{Pope_1997}].}
\label{fig:Water_absorption}
\end{figure}

An obvious question concerns the possibility of OH-absorption peaks in pure water. In first approximation water is remarkably transparent in the visible region. However, 
a detailed examination reveals the existence of absorption peaks, and that their de-excitations could be detected by a Bialkali photocathode. Figure~\ref{fig:Water_absorption} 
shows measurements of pure water absorption peaks by the authors listed in Ref. [\cite{Pope_1997}]. 
These absorption peaks are well understood in terms of molecular vibration modes of the water molecule. 
We conclude that the following wavelengths fall in the reach of a Bialkali photocathode: 742, 662, 605, 550, 514, 473, 449, 418 and 401~nm.

\begin{figure}[tbp]
\includegraphics[width=0.4\textwidth]{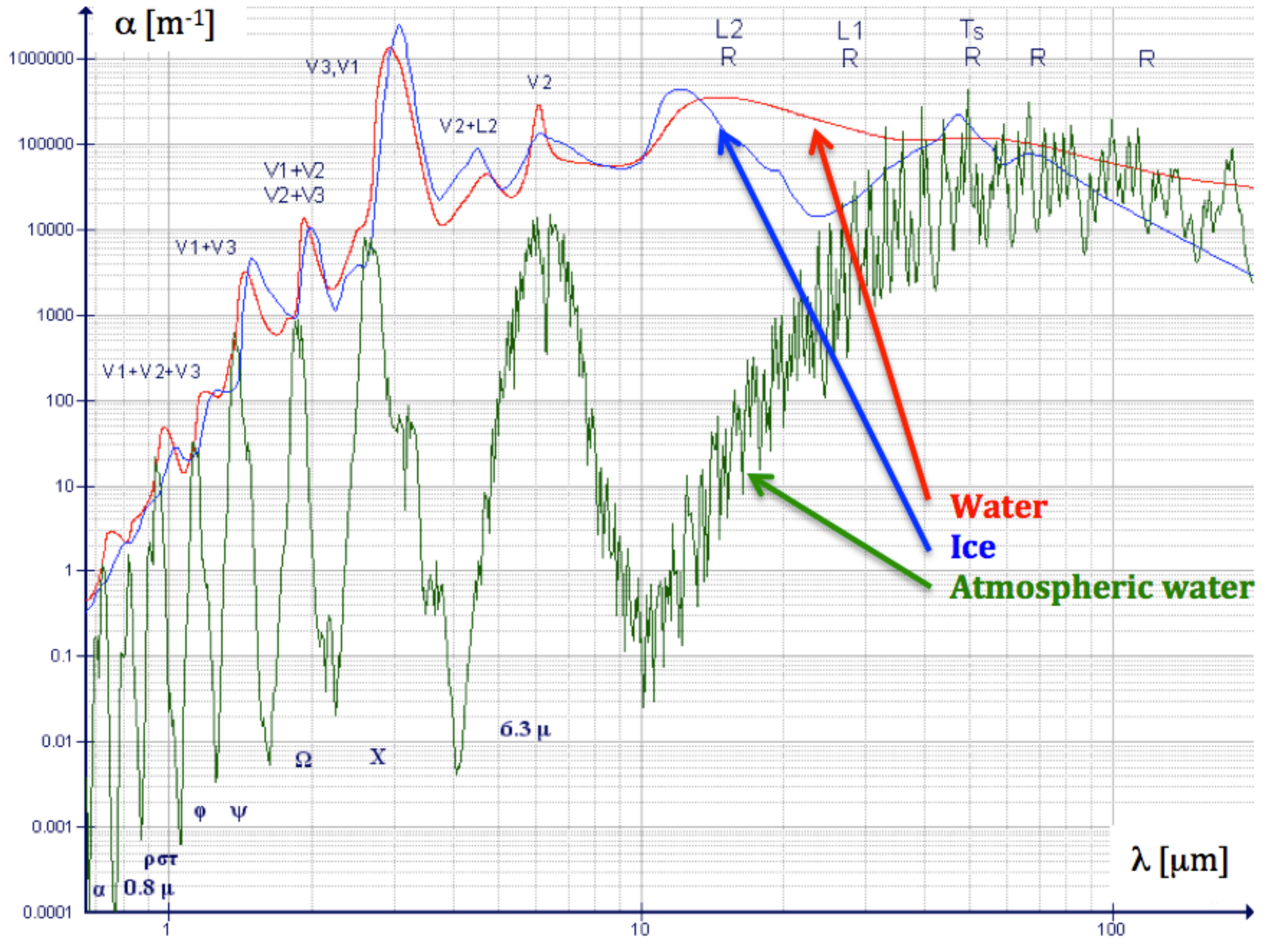}
\caption{Absorption spectrum (attenuation coefficient vs. wavelength) of liquid water, atmospheric water vapor and ice
between 667~nm and 2000~nm [\cite{Wozniak_2007}].}
\label{fig:Water_Ice_absorption}
\end{figure}

\begin{figure}[tbp]
\includegraphics[width=0.4\textwidth]{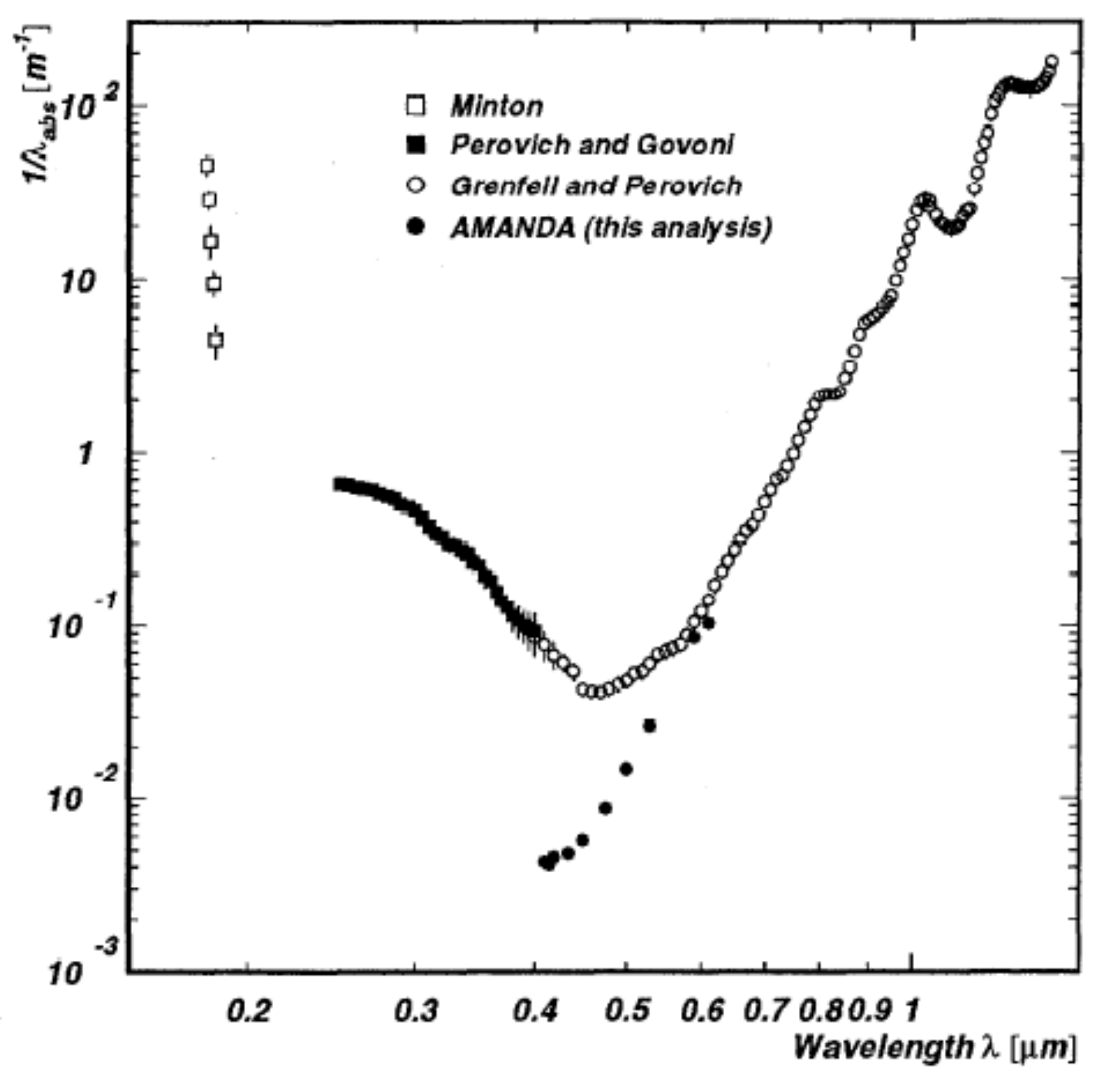}
\caption{Absorption coefficient of ice
between 250~nm and 2000~nm [\cite{Askebjer_1997}].}
\label{fig:Ice_absorption}
\end{figure}

Figures~\ref{fig:Water_Ice_absorption} and \ref{fig:Ice_absorption} show the photon absorption coefficient as a function of wavelength in liquid water, atmospheric water vapor 
and ice over a wide wavelength range [\cite{Pope_1997}],[\cite{Wozniak_2007}],[\cite{Askebjer_1997}]. Although absorption peaks are more clear at longer wavelengths, the medium 
transmission gets smaller. For example, the deep polar ice has a very long transmission in visible range ($\lambda_{Abs}$(450nm)~$\sim$150 meters), it is getting small at longer 
wavelengths ($\lambda_{Abs}$(600nm)~$\sim$10~m), and very small at IR wavelengths ($\lambda_{Abs}$(1000nm)~$\sim$0.1~m).

The molecular excitation by WIMPs might be different than photon excitation, as the WIMP interacts by collisions with nuclei. Closest particle to 
simulate the WIMP collisions are thermal neutrons. Typical spectroscopic experiments in this field measure a dominant 
peak $\sim$2.9~$\mu$m (peak is wavenumber 3400~cm$^{-1}$), located in the IR range.

It is clear that the efficiency to generate visible photons in water or ice by the Dark Matter is very small. A large radiator volume is required,
together with an extensive PMT coverage. Exmeriments such as the Ice Cube, the BaBar DIRC and the Super-Kamiokande experiment have a chance. 

In future, one may choose detectors working in the infrared wavelength region, packed 
10-20~cm apart of each other. It is essential that the detector operation is very stable to be able to see a very small modulation signal.

\section{OH-molecular vibrations in NaI(Tl) crystal}

Reference [\cite{Vavra_2014_a}] used proton recoils from hydrogen atoms to provide a possible new explanation of the DAMA signal. The signal was created 
through the scintillation mechanism of the NaI(Tl) crystal. The OH molecular vibrations discussed in this paper represent an additional 
mechanism which might result in a very small contribution to this signal, as the amount of OH-molecules is small in the NaI(Tl) crystal.

\section{More complicated molecules}

In principle, many oils might be used in the study of molecular vibrations. To analyze molecular vibrations in such molecules is much more complicated 
than to do so in di-atomic molecules. Such molecules have many more modes of vibration. However, if an experiment 
has a large volume of oil and single-photoelectron sensitivity, it may be sensitive to Dark Matter collisions, even with Bialkali photocathodes. 
Light emission from such molecules could be calibrated by very low energy neutrons. 

The KamLand experiment uses a mix of oil and scintillator. Therefore it may be sensitive to proton recoils leading to the scintillation, through 
a similar mechanism as the DAMA experiment. We have learned, that such an effort is under way in the KamLand 
experiment.\footnote[1]{J. Learned, Univ. of Hawaii, private communication.}

\section{Conclusion}

This paper provides a scheme within which molecular vibrations might play a role in the discovery of a light WIMP. We have described in detail
the vibrations of OH-molecules in Fused Silica and in pure water. A very small energy deposit at the level of 2-3~eV is needed to excite such vibrations, 
with the corresponding de-excitations detectable by a Bialkali photocathode. This is a factor of 2 lower in energy than is needed to excite
the classical scintillation mechanism in a NaI(Tl) crystal. Since the efficiency of the light emission via this process is small, a 
very large radiator volume of $e.g.$ water, Fused Silica or oil is required, together with extensive detector coverage with single-photon detection capability.

\section{Acknowledgements}

 I thank Prof. J. Learned for his enthusiastic support of this idea and suggestion to search for a DM signal in the KamLand experiment.

\end{document}